\documentclass[longnamesfirst,apjl]{emulateapj}
\usepackage{apjfonts,natbib,epsfig}
\tighten

\newcommand{\beq}{\begin{equation}}
\newcommand{\eeq}{\end{equation}}
\newcommand{\bea}{\begin{eqnarray}}
\newcommand{\eea}{\end{eqnarray}}

\begin{document}
\title{Cluster Merger Variance and  the Luminosity Gap Statistic }
\author{Milo\v s Milosavljevi\'c\altaffilmark{1,2}, 
Christopher J.~Miller\altaffilmark{3},
Steven R.~Furlanetto\altaffilmark{1}, and 
Asantha Cooray\altaffilmark{4}}
\altaffiltext{1}{Theoretical Astrophysics, Mail Code 130-33, California Institute of Technology, 1200 East California Boulevard, Pasadena, CA 91125.}
\altaffiltext{2}{Hubble Fellow.}
\altaffiltext{3}{Cerro-Tololo Inter-American Observatory, NOAO, Casilla 603, La Serena, Chile.}
\altaffiltext{4}{Department of Physics and Astronomy, University of California, Irvine, Irvine, CA 92617. }
\righthead{MILOSAVLJEVI\'C ET AL.}
\lefthead{THE LUMINOSITY GAP STATISTIC}

\begin{abstract}

The presence of multiple luminous galaxies in  clusters can be explained by the finite time over which a galaxy sinks to the center of the cluster and merges with the the central galaxy.  The simplest measurable statistic to quantify the dynamical age of a system of galaxies is the luminosity (magnitude) gap, which is the difference in photometric magnitude between the two most luminous galaxies.  We present a simple analytical estimate of the luminosity gap distribution in groups and clusters as a function of dark matter halo mass.  The luminosity gap is used to define ``fossil'' groups; we expect the fraction of fossil systems to exhibit a strong and model-independent trend with mass:   $\sim1$--$3\%$ of massive clusters and $\sim 5$--$40\%$ of groups should be fossil systems. We also show that, on cluster scales, the observed intrinsic scatter in the central galaxy luminosity-halo mass relation can be ascribed to dispersion in the merger histories of satellites within the cluster. We compare our predictions to the luminosity gap distribution in a sample of 730 clusters in the Sloan Digital Sky Survey C4 Catalog and find good agreement.   This suggests that  theoretical excursion set merger probabilities and the standard theory of dynamical segregation are valid  on cluster scales.

\keywords{ cosmology: observations --- dark matter --- galaxies: clusters: general --- galaxies: halos }

\end{abstract}

\section{Introduction}
\label{sec:intro}

\setcounter{footnote}{0}

Virialized cold dark matter halos grow hierarchically, by the merging of smaller virialized halos. Galaxies, which populate the  halos, also grow hierarchically by the merging of pre-existing galaxies. Halo merger rates can be estimated analytically using the excursion-set theory \citep{Bond:01}, which is commonly known as the Extended Press-Schechter formalism \citep{Press:74,Bower:91,Lacey:93}.  The merger rates can also be extracted from large-scale cosmological simulations. 
Direct measurement of merger rates in galaxy groups and clusters, however, is challenging.  While mergers can be identified by unrelaxed X-ray morphologies (e.g.,~\citealt{Jeltema:05}) or by the presence of radio halos produced by nonthermal particles associated with merger shocks (e.g., \citealt{Ensslin:02}), these methods are not yet precise enough to yield accurate estimates of merger rates.  We here show that a simple statistic, the luminosity gap, can be applied to  large galaxy surveys such as the Sloan Digital Sky Survey (SDSS) to test the predictions of the excursion-set merger probabilities.

Dark matter halos merge from the outside in. 
Mergers effectively begin when the components approach within a virial radius of each other; they conclude when the separate identities of the two halos have been erased, either by the merging of the halo centers or by the complete tidal disruption of the smaller halo.  The duration of the merger defined this way depends on the rate at which dynamical friction induces orbital decay of the tidally truncated subhalos. If a galaxy is located at the center of each subhalo, the galaxies appear as separate objects until the conclusion of the merger.  
The timescale of orbital decay can exceed the age of the system; this is why the number of galaxies in a groups or cluster (the ``occupation number'')  exceeds unity.

A confirmation of this paradigm can be found in the existence of ``fossil'' groups of galaxies, which contain a single ultraluminous galaxy at the center and no other galaxy brighter than $L_\star$.  The X-ray luminosity of fossil groups is comparable to that of rich clusters and indicates dynamical masses $\sim(10^{13}$--$10^{14}) \ {M}_\odot$. \citet{Jones:03} selected fossil groups with the criterion $\Delta{\rm mag}_{12}>2$, where the luminosity gap $\Delta{\rm mag}_{12}$ is defined as the difference in photometric magnitude between the most and second most luminous galaxies in the group; we adopt this definition as well.  Fossil groups have been identified as cluster-sized systems old enough that any previous merger with a halo hosting  a $\gtrsim L_\star$ galaxy has completed so that all luminous galaxies have agglomerated onto the central galaxy. Their incidence rate is $(8$--$20)\%$ among 
observed systems in this mass range, at least if the mass is inferred from the X-ray luminosity \citep{Jones:03}.  
Numerical simulations by \citet{Donghia:05} predict a larger fraction of fossil systems: $(33\pm16)\%$ for mass $10^{14}M_\odot$.  These investigations suggest that fossil groups, poor clusters, and rich clusters can be distinguished by the time elapsed since the last major merger.  We show here that the incidence rate of fossil groups can be estimated analytically from excursion set theory.

In \S~\ref{sec:dynamics}, we compute the time scale on which dynamical friction drives orbital inspiral of a tidally truncated subhalo inside the primary halo.  
In \S~\ref{sec:merger}, we estimate the subhalo mass distribution as a function of the subhalo mass and distance from the center of the primary halo.  In \S~\ref{sec:gap}, we calculate the luminosity distribution of the most luminous satellite galaxy within the cluster and the dependence of the luminosity gap on the halo mass.  We also derive the fraction of fossil systems in groups and clusters as a function of mass.  
In \S~\ref{sec:sdss}, we compare the predictions of our model to the luminosity gap distribution in 730 clusters from the SDSS C4 Catalog \citep{Miller:05}.  Finally, in  \S~\ref{sec:lcm_scatter}, we show that the observed intrinsic scatter in the $L_{\rm c}$--$M$ relation on cluster scales can be ascribed to dispersion in the merger histories. Throughout the paper we assume the standard cosmological model consistent with the results from the {\it Wilkinson Microwave Anisotropy Probe} \citep{Spergel:03}.

\section{Dynamical Evolution}
\label{sec:dynamics}

Consider a subhalo of mass $M_{\rm s}$ merging into a primary halo of mass $M_{\rm h}\geq M_{\rm s}$ at redshift $z_{\rm m}$. We define a ``merger'' as the time when the center of subhalo crosses the virial radius of the new composite halo of mass $M=M_{\rm h}+M_{\rm s}$.   After the halos coalesce, the subhalo spirals toward the center of the composite halo.  
The effective dynamical mass of the subhalo decreases after the merger because bound mass is tidally stripped as the orbit of the subhalo decays.  We denote the bound mass by $M_{\rm s}(R_{\rm s})$, where $R_{\rm s}$ is the tidal truncation radius.  This is related to the separation $r$ of the subhalo from the center of the composite halo via $M_{\rm s}(R_{\rm s})/R_{\rm s}^3=M(r)/r^3$, where $M(r)$ is the mass of the composite halo contained within radius $r$.  This relation yields $R_{\rm s}$ and $M_{\rm s}(R_{\rm s})$ in terms of $r$.

The subhalo experiences a torque $T=|{\bf r}\times {\bf F}|$, where ${\bf r}$ is its position relative to the center of the composite halo and ${\bf F}$ is force of dynamical friction (\citealt{Chandrasekhar:43}) $|{\bf F}|= 4\pi  G^2 M_{\rm s}(R_{\rm s})^2 \rho(r)\ln(\Lambda)/v^2$.  Here, $\rho(r)$ is the density of the composite halo at distance radius $r$, $\ln(\Lambda)$ is the Coulomb logarithm (which in principle depends on the orbit of the subhalo and on the orbital phase space distribution of dark matter), and $v$ is the velocity of the subhalo.  To simplify the calculations we assume circular orbits.  Then the velocity of the subhalo is the circular velocity in the composite halo $v=[GM(r)/r]^{1/2}$.  In numerical simulations of satellites in halos  $\ln(\Lambda) \sim 2$ \citep{Velazquez:99,Fellhauer:00}.  

The galaxies spiral toward the center of the composite halo at the rate
\beq
\label{eq:radius_deriv}
\frac{dr}{dt}=\frac{T}{M_{\rm s}(R_{\rm s})}\left(\frac{dJ}{dr}\right)^{-1} ,
\eeq
where $J=[G M(r) r]^{1/2}$ is the specific angular momentum associated with the orbit of the subhalo. 
We assume that the density profile of the halo follows the form of \citet{Navarro:97}, with the concentration parameter from the \citet{Bullock:01} fit. Equation (\ref{eq:radius_deriv}) can be integrated to find the distance of the subhalo from the center of the composite halo as a function of time.  

\section{Subhalo Distribution}
\label{sec:merger}

We would like the probability that a subhalo of mass $M_{\rm s}$ is located at distance $r$ from the center of the composite halo of mass $M$.  The extended Press-Schechter formalism does not directly yield such an expression because halos grow through an entire hierarchy of mergers.  We present a variation of one of the established models for the subhalo mass function (e.g.,~\citealt{Nusser:99,Fujita:02,Sheth:03,Lee:04,Oguri:04}).  These models ignore a number of issues, such as halo triaxiality and the evolution of substructure within substructure. Our confidence in their validity stems from their success in reproducing subhalo statistics in large-scale numerical simulations (e.g.,~\citealt{Zentner:05}), gravitational lensing observations (e.g.,~\citealt{Natarajan:04}), and the cluster luminosity function \citep{Cooray:05c}.  
Note that the known problems with the extended Press-Schechter merger rates are not severe for the mass ratios of interest \citep{Benson:05}.

\citet{Lacey:93} give an expression for the fraction of mass of a halo with mass $M$ at redshift $z$ that lies in progenitors with masses between $M_{\rm s}$ and $M_{\rm s}+dM_{\rm s}$ at redshift $z_{\rm m}$:
\bea
\label{eq:upcrossing}
f(M_{\rm s},z_{\rm m}|M,z)&=&\frac{\delta_{\rm c}(z_{\rm m})-\delta_{\rm c}(z)}{(2\pi)^{1/2}[\sigma^2(M_{\rm s})-\sigma^2(M)]^{3/2}}\left|\frac{d\sigma^2(M_{\rm s})}{dM_{\rm s}}\right|
\nonumber\\
&\times&\exp\left\{-\frac{[\delta_{\rm c}(z_{\rm m})-\delta_{\rm c}(z)]^2}{2[\sigma^2(M_{\rm s})-\sigma^2(M)]}\right\} ,
\eea
where $\sigma(M)$ is the mass variance on scale $M$ and $\delta_{\rm c}(z)$ is the critical overdensity   for collapse at redshift $z$. The progenitor mass function  is obtained by multiplying $f(M_{\rm s},z_{\rm m}|M,z)$ by the halo multiplicity factor, $dN/dM_{\rm s}=(M/M_{\rm s}) f$.

\citet{Lacey:93} define the formation redshift as the time at which the most massive progenitor halo contains half of the mass of the final halo.  Equation (\ref{eq:upcrossing}) is integrated with respect to $M_{\rm s}$ and differentiated with respect to $z_{\rm m}$ to obtain the growth rate of the fraction of the halo mass in large objects; we interpret this as the probability that a halo at redshift $z$ formed at a  redshift between  $z_{\rm m}$ and $z_{\rm m}+dz_{\rm m}$,
\bea
\label{eq:jump}
\frac{dP}{dz_{\rm m}}(z_{\rm m},M,z)&=& \frac{d}{dz_{\rm m}} \int_{M_{\rm min}}^{M} f (M_{\rm s},z_{\rm m}|M,z) dM_{\rm s} 
\eea 
where $M_{\rm min}\geq\onehalf M$ is a minimum mass cutoff.

Following \citet{Oguri:04}, we multiply the subhalo mass function by the formation redshift distribution 
\beq
\label{eq:subhalo_distribution}
\frac{d^2N}{dM_{\rm s}dz_{\rm m}}\sim \frac{dN}{dM_{\rm s}} \times\frac{dP}{dz_{\rm m}}
\eeq
and interpret the result as the probability that the halo acquired a subhalo of mass $M_{\rm s}$ at redshift $z_{\rm m}$.  This interpretation is only heuristic; doing better requires integrating over merger trees.
Such a treatment would improve upon the imprecise definition of ``formation time'' \citep{Cohn:05} and its likely correlation with subhalo properties, but we defer that to future work.  

The probability that a subhalo of mass $M_{\rm s}$ lies a distance $r$ from the halo center, as a function of the halo mass $M$ and redshift of observation $z$, is given by (we suppress the dependence on $M$ and $z$ everywhere)
\bea
\label{eq:dpdr1}
\frac{d^2N}{dM_{\rm s} dr}(M_{\rm s},r)&=&\frac{d^2N}{dM_{\rm s}dz_{\rm m}}(M_{\rm s},z_{\rm m})\frac{dz_{\rm m}}{dr} (M_{\rm s},z_{\rm m}) 
\eea 
for $r<R_{\rm vir}(M,z)$, where the average density of the halo is $200$ times the mean density of the universe within the virial radius, and $dz_{\rm m}/dr$ is the inverse of the derivative of the separation at redshift $z$ with respect to the merger redshift $z_{\rm m}$ (from eq.~[\ref{eq:radius_deriv}], after converting cosmic time to redshift). 

The luminosity gap is independent of the subhalo position, so we integrate equation (\ref{eq:dpdr1}) with respect to radius to obtain
\bea
\label{eq:mass_function}
\frac{dN_R}{dM_{\rm s}}(M_{\rm s})
&=& \int_{z_{\rm m}(M_{\rm s},R)}^{z_{\rm m}(M_{\rm s},0)} \frac{d^2N}{dM_{\rm s}dz_{\rm m}}(M_{\rm s},z_{\rm m}) dz_{\rm m} ,
\eea
where $R\leq R_{\rm vir}$ is a constant radius and $z_{\rm m}(M_{\rm s},r)$ is implicitly defined by the relation
\beq
\label{eq:zmerge_implicit}
R_{\rm vir}[z_{\rm m}(M_{\rm s},r)]-r=\int_z^{z_{\rm m}(M_{\rm s},r)} \frac{dr}{dz_{\rm m}'} dz_{\rm m}' .
\eeq
We solve (\ref{eq:zmerge_implicit}) approximately by evaluating the virial radius at $z$ rather than $z_{\rm m}$ and by evaluating $dr/dz_{\rm m}$ at $R_{\rm vir}(z)$.  The latter approximation is justified  because $dr/dz_{\rm m}$ depends on radius only weakly, especially when $r\sim R_{\rm vir}$.

The number of subhalos with mass greater than $M_{\rm s}$ is
\beq
\label{eq:number_cumulative}
N_R(M_{\rm s})=\int_{M_{\rm s}}^{M} \frac{dN_R}{dM'}(M') dM' .
\eeq
Equation (\ref{eq:number_cumulative}) defines a one-to-one relation between $M_s$ of a particular subhalo and the expected number of subhalos above this threshold.  It is straightforward to show that the distribution of the most massive surviving subhalo mass, $M_2$, is
\beq
\label{eq:distribution_max_subhalo}
\frac{dN}{dM_2}= -\frac{d}{dM_2}e^{-N_R(M_2)} .
\eeq
Similarly, the distribution of the second most massive surviving subhalo mass, $M_3$, is
\beq
\label{eq:distribution_sec_subhalo}
\frac{dN}{dM_3}= -\frac{d}{dM_3} \{[1+N_R(M_3)]e^{-N_R(M_3)}\} .
\eeq

\begin{figure}
\epsscale{1.2}
\plotone{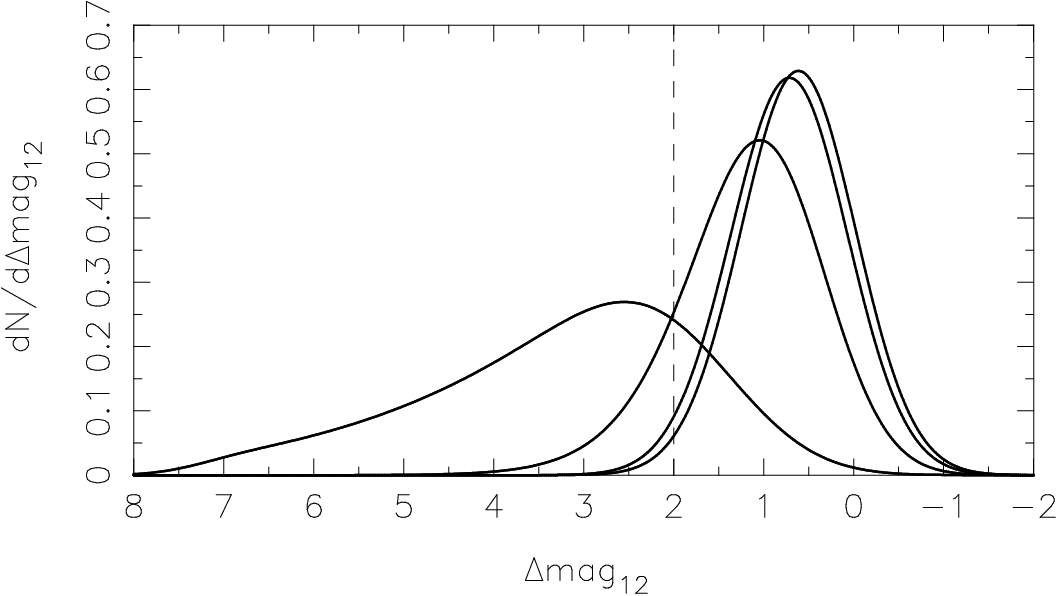}
\caption{Probability distribution for the luminosity gap $\Delta{\rm mag}_{12}$ at $z=0$  calculated using equation (\ref{eq:nmag}).  From left to right, the curves correspond to halos with mass $M=(10^{12.5}, \, 10^{13.5}, \, 10^{14.5}, \, 10^{15.5}) \ {M}_\odot$.  We assumed $\ln\Lambda=2$, $M_{\rm min}=\onehalf M$, and $R= R_{\rm vir}(M)$ throughout. While the initial mass of the satellite is always smaller than the mass of the primary, the luminosity of the satellite galaxy may exceed that of the primary because of the scatter in the $L_{\rm c}$--$M$ relation (equation \ref{eq:lcm_scatter}), leading to a negative luminosity gap; 
the measured quantity is $|\Delta{\rm mag}_{12}|$. 
\label{fig:nmag}}
\end{figure}

\section{The Luminosity Gap Distribution}
\label{sec:gap}

We next calculate the distribution of luminosities of the first and second most luminous  satellites in a galaxy cluster.  Thus we need to relate the initial mass of a subhalo to the luminosity $L_{\rm c}$ of its central galaxy. 
At $z \approx 0$, $L_{\rm c}$ in any halo is tightly correlated with its mass \citep{Vale:04,Cooray:05a,Cooray:05b}, with a functional form
\beq
\label{eq:lcm}
L_{\rm c}(M)=L_0 \frac{(M/M_0)^a}{[b+(M/M_0)^{cd}]^{1/d}} ,
\eeq
Our fit to the $r$-band luminosities in \citet{Seljak:05} yields $L_0=5.7\times10^{9}L_\odot$, $M_0=2\times10^{11} M_\odot$, $a=4$, $b=0.57$, $c=3.78$, and $d=0.23$. The  relation possesses lognormal intrinsic scatter \citep{Cooray:05b}
\beq
\label{eq:lcm_scatter}
\frac{dN}{dL}(L|M)= \frac{1}{\sqrt{2\pi} \ln(10)\Sigma L}
\exp\left\{-\frac{\log[L/L_{\rm c}(M)]^2}{2\Sigma^2}\right\} ,
\eeq
where in the $r$-band  $\Sigma\approx 0.17^{+0.02}_{-0.01}$ in clusters \citep{Cooray:05d}.

We assume that the most luminous galaxy is located at the center of the composite halo.  Then the remaining subhalos cannot contain the most luminous galaxy in the cluster; the most luminous galaxy in a subhalo is the second most luminous member of the cluster.  The distribution of galaxy luminosities associated with the $k$th most massive subhalo is then
\beq
\label{eq:dndl_k}
\frac{dN}{dL_k} =\int_0^M  \frac{dN}{dL} (L_k|M_k)\frac{dN}{dM_k}  dM_k .
\eeq

The luminosity gap $\Delta{\rm mag}_{1k}$ is the (observed) magnitude difference between the first and the  $k$th most luminous galaxies in a cluster.   The most luminous galaxy is the central galaxy, and the 
second one lies in the largest surviving subhalo. In systems of mass $M$ the gap is distributed as
\bea
\label{eq:nmag}
& &\frac{dN}{d\Delta{\rm mag}_{1k}}(\Delta{\rm mag}_{1k}|M)=
\int_0^\infty \int_0^\infty \frac{dN}{dL_k}(L_k,M)\frac{dN}{dL} (L|M) \nonumber\\& &\ \ \ \ \ \ \ \ \ \times 
 \delta\left[\Delta{\rm mag}_{1k}+\frac{5}{2}\log\left(\frac{L_k}{L}\right)\right] dL_kdL ,
\eea
where $\delta(x)$ is the Dirac delta-function. In Figure \ref{fig:nmag} we plot $dN/d\Delta{\rm mag}_{12}$ for halos of various masses; the median luminosity gap is larger in smaller halos.

\begin{figure}
\epsscale{1.2}
\plotone{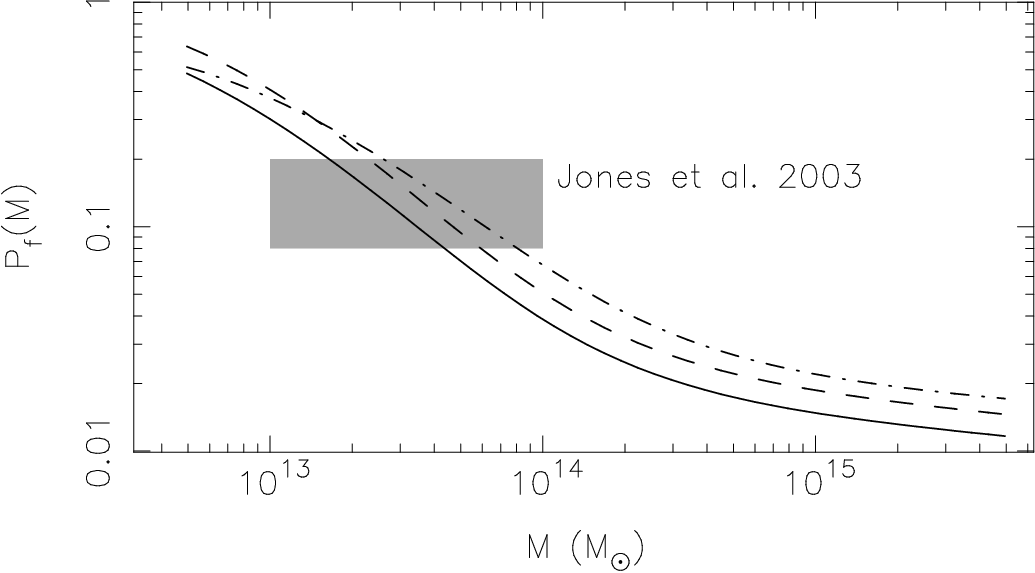}
\caption{Probability $P_{\rm f}(M)$ that a halo of mass $M$ contains a fossil system of galaxies. 
The curves assume $\ln\Lambda=1$ and $M_{\rm min}=\onehalf M$ ({\it solid line}), $\ln\Lambda=2$ and $M_{\rm min}=\onehalf M$ ({\it dashed line}), and  $\ln\Lambda=1$ and $M_{\rm min}=\frac{3}{4} M$ ({\it dot-dashed line}).  The shaded rectangle is the measurement of \citet{Jones:03}. \label{fig:fossil}}
\end{figure}

The incidence rate $P_{\rm f}(M)$ of ``fossil'' systems ($\Delta{\rm mag}_{12}>2$)  can then be calculated by integrating equation (\ref{eq:nmag}).
In Figure \ref{fig:fossil} we plot $P_{\rm f}(M)$ for several combinations of input parameters.  On scales $M\sim10^{14}M_\odot$, the probability that the system is fossil is $\sim (3$--$6)\%$, fairly close to the observed value \citep{Jones:03}.  It is, however, smaller than the same probability 
estimated from the simulations of \citet[see \S~\ref{sec:intro}]{Donghia:05}.
The three curves illustrate the errors we expect from our simplified treatment; they do not affect our qualitative results but may be distinguishable with more extensive observations.

\section{The Luminosity Gap in SDSS-C4 Clusters}
\label{sec:sdss}

We measured the luminosity gap distribution on 730 clusters in the SDSS C4 Cluster Catalog \citep{Miller:05} at mean redshift $\langle z\rangle=0.087$.  
The three brightest cluster members were identified from the SDSS photometry
as being within a projected $500h^{-1}\textrm{ kpc}$ radius of the center of the cluster.  Additionally, these galaxies must have $m_r-m_i$ colors that lie within $2\sigma$ of the E/S0 cluster ridgeline as determined by spectroscopically confirmed members \citep{Visvanathan:77}.  We utilize extinction corrected model-fit magnitudes and apply $z=0$ $K$-corrections (version 3.2, \citealt{Blanton:03}).

\begin{figure}
\epsscale{1.2}
\plotone{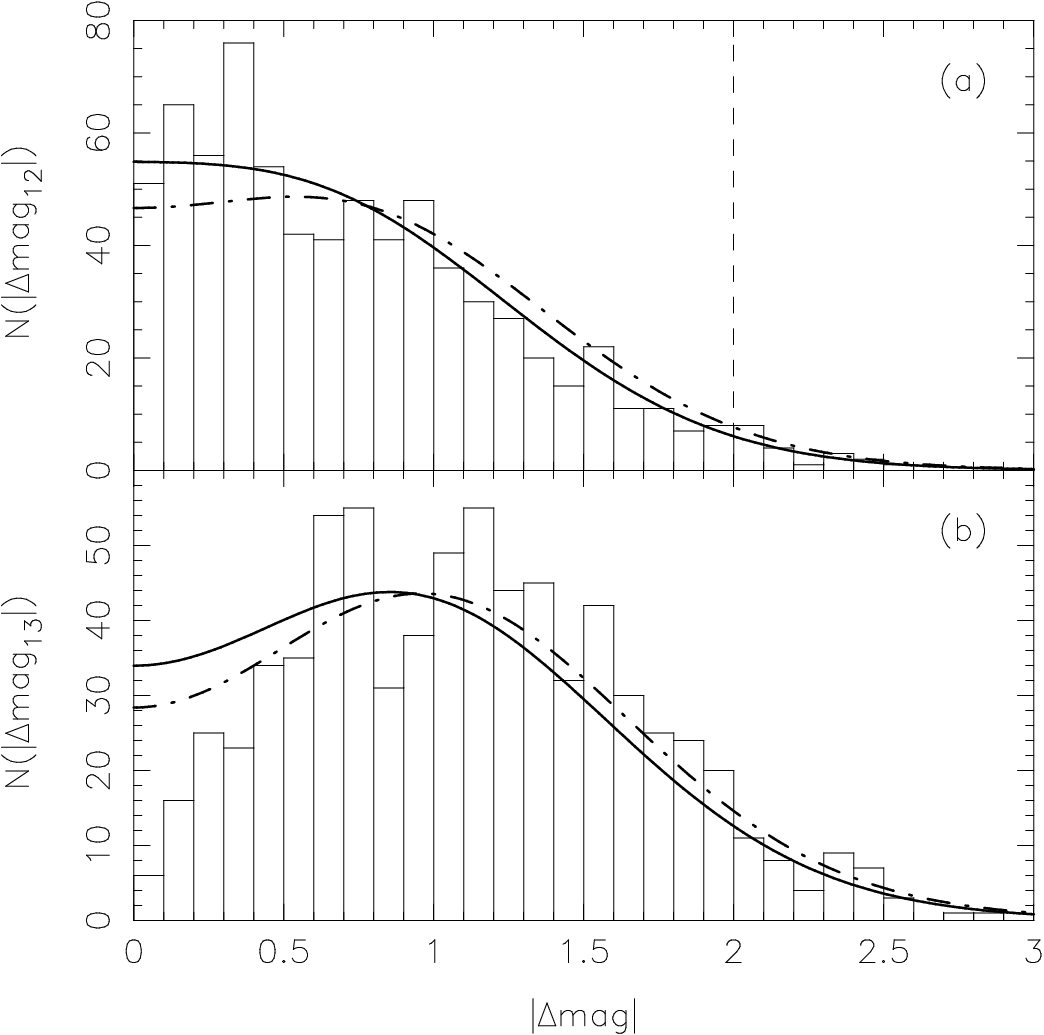}
\caption{$r$-band luminosity gap distribution from 730 clusters in the SDSS C4 Catalog \citep{Miller:05}. We evaluate the luminosity gap relative to, \emph{(a)} the first and second most luminous galaxies, and \emph{(b)} the first and the third most luminous galaxies.  We also show predictions of our model with $\ln\Lambda=1$ and $M_{\rm min}=\frac{1}{2}M$ ({\it thick line}) and with $\ln\Lambda=1$ and $M_{\rm min}=\frac{1}{2}M$ ({\it dot-dashed line}).  Fossil systems are located to the right of the dashed line; $P_{\rm f} \approx 2.9\%$ in the data and $2.7\%$ in the model.\label{fig:hist}}
\end{figure}

The masses of these clusters are estimated from the total $r$-band luminosities via $\log (h^{-1}M)\approx-2.46+1.45\log (h^{-2}L_r)$; $95\%$ lie in the range $(0.5$--$10)\times10^{14}(h/0.7)^{-1} \ {M}_\odot$. The total luminosity is a better mass estimator than the line-of-sight velocity dispersion or the richness of the cluster (\citealt{Miller:05}; see also \citealt{LinMohrStanford:04}, \citealt{Popesso:05}, and references therein). A histogram of the luminosity gap distribution is shown in Figure \ref{fig:hist}. The mean luminosity gap is $\langle\Delta{\rm mag}_{12}\rangle\approx0.75$.

To predict the luminosity gap distribution of the C4 sample, we multiply $dN/d\Delta{\rm mag}_{12}$ of equation (\ref{eq:nmag}) at $z=0$ by the mass function of C4 clusters and integrate over mass.  The 
resulting composite model luminosity gap distribution is shown by the thick solid and dot-dashed lines in Figure \ref{fig:hist}\emph{a}.  The agreement of the model and data is remarkable for the choice of the dynamical friction parameter $\ln\Lambda=1$.  For $\ln\Lambda\gtrsim 2$, the model distribution develops a local minimum at $\Delta{\rm mag}_{12}=0$ and its maximum shifts to $\Delta{\rm mag}_{12}\sim 0.6$ (dot-dashed line in Fig. \ref{fig:hist}).  Such behavior is not apparent in C4 clusters, but it has been detected 
in luminous red galaxies (LRG) in SDSS imaging data by \citet{Loh:05}, who measured the luminosity gap within $1.0h^{-1}\textrm{ Mpc}$ of LRGs and found that the peak moved to positive values in underdense fields.
These curves illustrate the sensitivity of the luminosity gap to the detailed properties of mergers; clearly the qualitative fit is excellent, but more detailed model-fitting may in the future enable  tests of particular aspects of merger dynamics.

In Figure \ref{fig:hist}\emph{b} we  present the luminosity gap distribution relative to the third most luminous galaxy in the cluster (see eq. [\ref{eq:distribution_sec_subhalo}]).
The model overpredicts the frequency of small $\Delta{\rm mag}_{13}$ 
because it treats the luminosities of the second and third most luminous galaxy as independent, whereas the former must exceed the latter by definition.

\section{The Origin of the $L_{\rm c}$--$M$ Scatter}
\label{sec:lcm_scatter}

The final accretion of a satellite halo at the center of the primary  halo is accompanied by an instantaneous increase of the central galaxy's luminosity.  Therefore, the accretion history is a source of intrinsic scatter in the $L_{\rm c}$--$M$ relation.   
We estimate the dispersion $\Sigma$ of the  central galaxy luminosity arising through the merger variance alone as the average luminosity of the most massive satellite galaxy, $\langle L_2(M) \rangle$,  assuming no intrinsic scatter in the mass-luminosity relation, $\Sigma (M) \sim \langle L_2(M)\rangle/\ln(10)L_{\rm c}(M)$.\footnote{A more accurate expression, to be presented in a subsequent paper, involves a sum over the variances of $L_k(M)$ in the place of $\langle L_2(M)\rangle$.}  In our model,
\beq
\langle L_2(M)\rangle  = \int_0^M L_{\rm c}(M_2) \frac{dN}{dM_2}(M,M_2)dM_2 .
\eeq

We average the dispersion calculated this way over the mass function of clusters in our C4 sample; it is $\bar\Sigma=0.23\pm0.01$, depending on the choice of $\ln\Lambda$.  This is not far from our adopted value $\Sigma_r=0.17$ and from $\Sigma\approx0.168$ measured by \citet{Yang:03}.  Therefore, the dispersion in the $L_{\rm c}$--$M$ relation in clusters can be explained by the dispersion in accretion histories. If this interpretation is correct, $\Delta{\rm mag}_{1k}$ at fixed cluster mass would be correlated with the central galaxy luminosity, implying that the factors of $dN/dL(L|M)$ in equations (\ref{eq:dndl_k}) and (\ref{eq:nmag}) must be replaced by a single multivariate distribution in $(L,L_k)$;
 we defer this possibility to a more detailed treatment.

\acknowledgements

We are grateful to D.~Buote, C.~Conselice, E.~Pierpaoli, C.~Sarazin, and J.~Taylor for inspiring discussions.  This work was supported at Caltech by a postdoctoral fellowship to M.~M.\ from the Sherman Fairchild Foundation and Hubble Fellowship grant HST-HF-01188.01-A awarded by the Space Telescope Science Institute, which is operated by the Association of Universities for Research in Astronomy, Inc., for NASA under contract NAS5-26555.  The research has made use of data obtained from or software provided by the US National Virtual Observatory, which is sponsored by the National Science Foundation.

\end{document}